\documentclass[epj]{svjour}

\usepackage{amsmath}
\usepackage{amsfonts}
\usepackage{amssymb}
\usepackage{graphicx}
\usepackage{cite}

\begin{document}

\title{On magnetic susceptibility of a spin-S impurity
in nearly ferromagnetic Fermi liquid}

\author{I.\,S.\,Burmistrov\inst{1,2,3}\thanks{e-mail: burmi@itp.ac.ru}
} \institute{L.\,D.\,Landau Institute for Theoretical Physics RAS,
117940 Moscow, Russia\and Institute for Theoretical Physics,
University of Amsterdam, 1018XE Amsterdam, The Netherlands\and
Department of Theoretical Physics, Moscow Institute of Physics and
Technology, 141700 Moscow, Russia}

\date{}

\abstract{ We present the renormalization group analysis for the
problem of a spin-S impurity in nearly ferromagnetic Fermi liquid.
We evaluate the renormalization group function that governs the
temperature behavior of the invariant charge to the second order
of both weak and strong coupling expansions. It allows us to
determine behavior of the zero field magnetic susceptibility of
impurity at low and high temperatures. We predict that derivative
of the susceptibility with temperature should always have the
maximum. \PACS{{75.20.-g}{Diamagnetism, paramagnetism, and
superparamagnetism}
\and {75.20.Hr}{Local moments in compounds and alloys; Kondo
effect, valence fluctuations, heavy fermions}
\and {75.20.En}{Metals and alloys}} } \maketitle

\section{Introduction}

The study of droplets of a local order in a non-ordered  phase
remains on a frontier of the modern condensed matter physics.
Especially interesting situation appears when the non-ordered
phase is near criticality~\cite{Criticality}. A particular example
of it is the problem of a magnetic impurity in Fermi liquid close
to the ferromagnetic instability~\cite{Stoner} which bares on the
same fundamental issue as the Kondo effect~\cite{Kondo,KondoBA}
but remains only partially solved so far. The problem was
pioneered by Larkin and Melnikov~\cite{LarkinMelnikov} who studied
a magnetic impurity in nearly ferromagnetic Fermi liquid, i.e., in
Fermi liquid with $1+F_0^\sigma \ll 1$ where $F_0^\sigma$ denotes
the standard Landau coefficient in the triplet channel~\cite{AGD}.
It was shown that a magnetic impurity acquires a giant magnetic
moment by inducing a droplet of spin-polarized electrons with a
size $a (1+F_0^\sigma)^{-1/2}$ which greatly exceeds length scale
$a$ which is of the order of an interatomic distance. The
theoretical description of the phenomenon is based on the concept
of paramagnons~\cite{DoniachEngelsberg} which are the low-energy
collective excitations in Fermi liquid close to the ferromagnetic
instability.

Experimentally the impurities with giant magnetic moments in
nearly ferromagnetic host metals have been intensively studying
since early sixties~\cite{EarlySixties}. The systems involve Fe
dissolved in metal alloys, e.g., Ni$_3$Ga with $F_0^\sigma=-0.97$,
Ni in Pd host ($F_0^\sigma=-0.9$) and Co impurities in Pt
($F_0^\sigma=-0.5 $) ~\cite{KorenblitShender}. Of late years the
giant moment of Co impurities has been experimentally observed in
alkali metals (Cs and Na) where it has been attributed to the
possible instability in alkali metals towards formation of a
charge density wave~\cite{Bergmann}. The effect of paramagnons has
been also studied in liquid He$^3$ which is another example of
Fermi liquid not so far from the ferromagnetic instability
($F_0^\sigma=-0.66$)~\cite{He}.

In the paper we consider a dilute system of magnetic impurities
with density $n_\textrm{imp}\ll a^{-3}(1+F_0^\sigma)^{3/2}$ in
nearly ferromagnetic Fermi liquid such that we do not need to take
into account the interaction between impurity spins. We present
the detailed renormalization group analysis for the problem both
in the weak coupling limit which first was considered in the
seminal paper~\cite{LarkinMelnikov} by Larkin and Melnikov and in
the strong coupling limit. We investigate the temperature ($T$)
behavior of the zero field magnetic susceptibility $\chi(T)$ of
the impurity. We find that the function $T \chi(T)$ decreases as
temperature is lowered and determine its asymptotics at low and
high temperatures (cf. Eqs.~\eqref{chiTWF} and \eqref{chiTSF}). We
show that derivative of the function $T\chi(T)$ with logarithm of
the temperature has a maximum.

We start out, in Sec.~\ref{Model} from the formulation of the
effective low-energy description for the imaginary time dynamics
of a single spin-S impurity in nearly ferromagnetic Fermi liquid.
In Sec.~\ref{RGA} we analyze the effect of the interaction with
paramagnons in the first and second orders of the weak coupling
expansion. In Sec.~\ref{RGS} we perform strong coupling analysis
of the interaction with paramagnons. On the basis of the
perturbative results we derive the renormalization group equation
that governs the change of the effective coupling with temperature
and present results for its temperature dependence
(Sec.~\ref{RNT}). In Sec.~\ref{Suscept} we derive the asymptotics
for the zero field magnetic susceptibility of impurity at low and
high temperatures. We end the paper with conclusions
(Sec.~\ref{Conc}).

\section{Model\label{Model}}

A single spin-S impurity placed into Fermi liquid at point
$\mathbf{r}=0$ is described by the total hamiltonian
$H=H_\textrm{el}+H_\textrm{ex}$ where $H_\textrm{el}$ denotes the
hamiltonian of electrons and
\begin{equation}
H_\textrm{ex} = J S^\alpha \psi^\dag(0)\sigma^\alpha
\psi(0)\label{HexIn}
\end{equation}
is the exchange hamiltonian. Here $\psi^\dag(\mathbf{r})$ and
$\psi(\mathbf{r})$ are the creation and annihilation operators for
electrons and $\sigma^\alpha$ with $\alpha=x,y,z$ denotes the
Pauli matrices. In the paper we consider the case of small
exchange coupling and nearly ferromagnetic Fermi liquid such that
the conditions $1~+~F_0^\sigma\ll |J|\nu \ll 1$ with $\nu$ being
the electron density of states per one spin are satisfied.
 Integrating out the electrons to the lowest
order in the parameter $(J\nu)^2\ll 1$ we obtain the effective
action for the impurity spin dynamics in the imaginary
time~\cite{LarkinMelnikov}
\begin{equation}
\mathcal{S}_\textrm{eff} = \frac{J^2}{2}\int_0^\beta d\tau_1
d\tau_2 \mathcal{D}(0,\tau_1-\tau_2) \mathbf{S}(\tau_1)
\mathbf{S}(\tau_2).\label{Seff}
\end{equation}
Here $\beta =1/T$ and the kernel $\mathcal{D}(\mathbf{r},\tau)$ of
the effective action~\eqref{Seff} is the spin-spin correlation
function of electrons
\begin{equation}
\mathcal{D}(\mathbf{r},\tau)\delta^{\alpha\beta} = \langle
\psi^\dag_b(0,0)\sigma^\alpha_{bc} \psi_c(0,0)
\psi^\dag_d(\mathbf{r},\tau)\sigma^\beta_{df}
\psi_f(\mathbf{r},\tau)\rangle .
\end{equation}
Assuming that electrons can be described in terms of Fermi liquid
close to the ferromagnetic instability then the spin-spin
correlation function $\mathcal{D}(\mathbf{k},\omega_n)$ strongly
depends on wave vector $\mathbf{k}$ and frequency $\omega_n=2\pi T
n$ even if they are small $|\omega_n|\ll k v_F$ and $k\ll
p_F$~\cite{DoniachEngelsberg,LarkinMelnikov}
\begin{equation}
\mathcal{D}(\mathbf{k},\omega_n) = \frac{2\nu}{1+F_0^\sigma+a^2
k^2 + \pi |\omega_n|/(2k v_F)}. \label{D1}
\end{equation}
Here $\nu=p_F^2/(2\pi^2 v_F)$ with $v_F$ and $p_F$ being the Fermi
velocity and the Fermi momentum respectively. The scale $a$ cannot
be expressed in terms of the Fermi liquid parameters. In the
ladder approximation $a$ was found to equal
$p_F^{-1}/\sqrt{12}$~\cite{DoniachEngelsberg}. However, this
approximation cannot be justified in nearly ferromagnetic Fermi
liquid. For example, estimate from the experimental data on
He$^{3}$ yields $a\approx 0.4 p_F^{-1}$~\cite{He}.

The function $\mathcal{D}(\mathbf{k},\omega_n)$ describes the
propagation of low energy boson excitations usually called
paramagnons. For the reasons to be explained shortly we introduce
\begin{eqnarray}
D(\omega_n) &=& \int \frac{d^3 k}{(2\pi)^3}\left [
\mathcal{D}(\mathbf{k},\omega_n)-\mathcal{D}(\mathbf{k},0)\right
]\notag \\
&=& -\frac{\pi \nu^2|\omega_n|}{2a^2p_F^2(1+F_0^\sigma)}
\mathrm{D}\left(\frac{|\omega_n|}{E_0}\right)\label{Domega}
\end{eqnarray}
where $\mathrm{D}(0)=1$ and $\mathrm{D}(z)\to 4\pi
/(3^{3/2}z^{2/3})$ in the limit $z\to\infty$~\cite{Footnote2}. As
it was first shown in Ref.~\cite{LarkinMelnikov} the contribution
from high frequencies $|\omega_n|\gtrsim
E_0=4E_F(1+F_0^\sigma)^{3/2}/(\pi a p_F)$ with $E_F$ being the
Fermi energy does not lead to any divergences. In what follows we
substitute $\mathcal{D}(0,\tau)$ in the effective
action~\eqref{Seff} by $D(\tau)=T\sum_{\omega_n} D(\omega_n)
e^{-i\omega_n \tau}$ where $D(\omega_n)$ is given by
Eq.~\eqref{Domega} with $\mathrm{D}(z) =\Theta(1-z)$. Here
$\Theta(z)$ denotes the Heaviside step function.

In general, the physics associated with hamiltonian~\eqref{HexIn}
strongly depends on sign of the exchange coupling $J$. However, as
one can see from effective action~\eqref{Seff}, the effect of
paramagnons is independent of $\textrm{sgn}\, J$. We discuss this
point in Sec.~\ref{Conc} below.

\section{Renormalization group analysis in the weak coupling limit\label{RGA}}

Equations~\eqref{Seff} and \eqref{Domega} indicate that there is a
single dimensionless parameter
\begin{equation}
g_0 = \frac{J^2\nu^2}{(ap_F)^2(1+F_0^\sigma)}\label{defg0}
\end{equation}
that governs the low-energy dynamics of the impurity spin. In the
weak coupling limit in which we assume $g_0$ being small it is
convenient to transform the effective action~\eqref{Seff} as
follows i) to decouple the spin-spin interaction by a paramagnon
field $\lambda_\alpha$ with a help of the Hubbard-Stratonovich
transformation~\cite{HS}; ii) to introduce the Abri\-kosov
pseudofermions with creation $c^\dag_m$ and annihilation $c_m$
operators in which a subscript $m$ runs from $-S$ to
$S$~\cite{Abrikosov,IzyumovSkryabin}. Then the effective action
becomes
\begin{eqnarray}
\mathcal{S}_\textrm{eff} &=& \int_0^\beta d\tau
c^\dag_{m}(\tau)\left [ (\partial_\tau +\eta)\delta_{mm^\prime} -
J S^\alpha_{mm^\prime} \lambda_\alpha(\tau)\right ]
c_{m^\prime}(\tau)\notag
\\
&-&\frac{1}{2} \int_0^\beta d\tau_1d\tau_2 \lambda_\alpha(\tau_1)
D^{-1}(\tau_1-\tau_2)\lambda_\alpha(\tau_2)\label{SeffWC}
\end{eqnarray}
In order to eliminate the contributions to the physical quantities
from nonphysical states we have introduced the chemical potential
$\eta$ which we should tend to minus infinity, $\eta\to -\infty$.
At the end of all calculations we leave only the leading order
term in series expansion of a physical quantity in powers of
$\exp(\beta \eta)$~\cite{Abrikosov,IzyumovSkryabin}. This
procedure results in the absence of renormalization for the
paramagnon propagator $D(\omega_n)$.

The diagrammatic technique for the action~\eqref{SeffWC} involves
the paramagnon propagator $D(\omega_n)$, the pseudofermion
propagator $G_{mm^\prime}(\epsilon_n)$  where $\epsilon_n= \pi T
(2n+1)$ and the para\-magnon-pseudofermion interaction vertex
$\Gamma_S(\omega_n,\epsilon_n) S^\alpha_{mm^\prime}$. In the zero
order approximation they are given as
\begin{equation}
G_{mm^\prime}(\epsilon_n)=G(\epsilon_n)\delta_{mm^\prime} =
\frac{\delta_{mm^\prime}}{i\epsilon_n+\eta},\qquad
\Gamma_S(\omega_n,\epsilon_n) = 1 . \label{bare}
\end{equation}
By using the standard diagrammatic technique we create the
perturbative expansion for action~\eqref{SeffWC} in powers of
$g_0$. For a sake of simplicity we shall analyze the action at
zero temperature. Therefore at the end of all calculations the
analytic continuation from upper half complex plane to the real
axis $i\epsilon_n\to\epsilon+i 0$ and $i\omega_n\to \omega +i 0$
will be performed. The basic objects to consider are the
pseudofermion self energy $\Sigma(E) = E-G^{-1}(E)$ and vertex
$\Gamma_s(\omega=0,E)$ where we introduce $E=\epsilon+\eta$ for a
brevity.

The theory described by action~\eqref{SeffWC} remains invariant
under the set of renormalization group transformations $\Gamma\to
z_1 \Gamma$, $\Gamma_S\to z_2 \Gamma_S$ and $J\nu\to z_1 z_2^{-1}
J\nu$ where $\Gamma(E) = 1-\Sigma(E)/E$~\cite{BogolubovShirkov}.
It allows us to construct the so-called invariant charge
\begin{equation}
g(E) = g_0 \frac{\Gamma_S^2(0,E)}{\Gamma^2(E)} \label{inv}
\end{equation}
that remains invariant under the above transformations. Now we
shall compute the functions $\Gamma_S(0,E)$ and $\Gamma(E)$
perturbatively in $g_0$ upto the second order~\cite{Maleev}. We
notice that we are interested only in terms which involve powers
of large logarithms $\ln E_0/|E|$. Since they appear only in the
real parts of $\Gamma_S(0,E)$ and $\Sigma(E)$ we shall omit
imaginary parts of them in the final results presented below.

\begin{figure}[tbp]\sidecaption
\includegraphics[width=30mm]{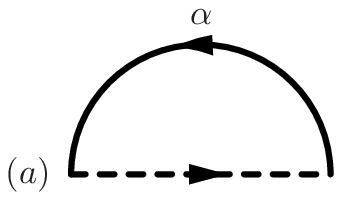}
\hspace{0.5cm}\includegraphics[width=30mm]{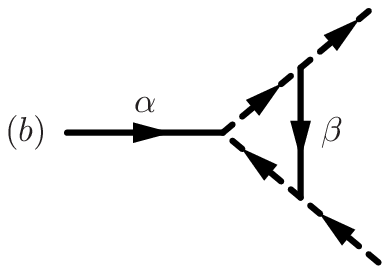} \caption{The
first order contributions to (a) the self energy
$\Sigma(\epsilon_n)$, (b) the vertex
$\Gamma_s(\omega_n,\epsilon_n) S^\alpha$. Solid line denotes the
paramagnon propagator $D(\omega_n)$ whereas dashed line stands for
pseudofermion propagator $G(\epsilon_n)$.} \label{Fig1}
\end{figure}

\subsection{First order contributions}

The corrections to the self energy $\Sigma(\epsilon_n)$ and vertex
function $\Gamma_S(\omega_n,\epsilon_n)$ of first order in $g_0$
are (see Fig.~\ref{Fig1})
\begin{eqnarray}
\Sigma^{(1)}(\epsilon_n) &=& J^2 \mathbf{S}^2_0 T\sum_{\omega_s}
D(\omega_s)G(\omega_s+\epsilon_n),\label{sigma1}\\
\Gamma_S^{(1)}(\omega_n,\epsilon_n) &=& J^2 \mathbf{S}^2_1
T\sum_{\omega_s} D(\omega_s)G(\omega_s+\epsilon_n)\notag \\
&&\hspace{1.3cm}\times G(\omega_s+\epsilon_n+\omega_n),
\label{vertex1}
\end{eqnarray}
where $\mathbf{S}^2_a = S(S+1)-a$. Evaluation of
expressions~\eqref{sigma1} and \eqref{vertex1} at vanishing
temperature yields
\begin{eqnarray}
\Sigma^{(1)}(E) = &-&\frac{g_0}{2}\mathbf{S}^2_0 E \ln \frac{E_0}{|E|},\label{Fsigma1}\\
\Gamma_S^{(1)}(0,E) =&&\frac{g_0}{2}\mathbf{S}^2_1 \ln
\frac{E_0}{|E|}.\label{Fvertex1}
\end{eqnarray}

\begin{figure}[tbp]
\includegraphics[width=30mm]{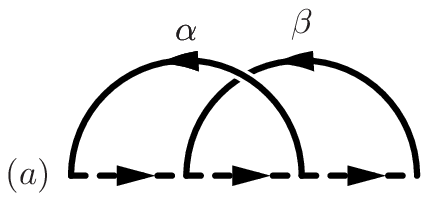}
\hspace{0.5cm}\includegraphics[width=30mm]{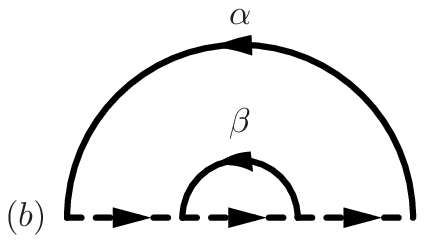}
\caption{The
second order contributions to the self energy
$\Sigma(\epsilon_n)$} \label{Fig2}
\end{figure}

\subsection{Second order contributions}

\begin{figure}[tbp]
\includegraphics[width=30mm]{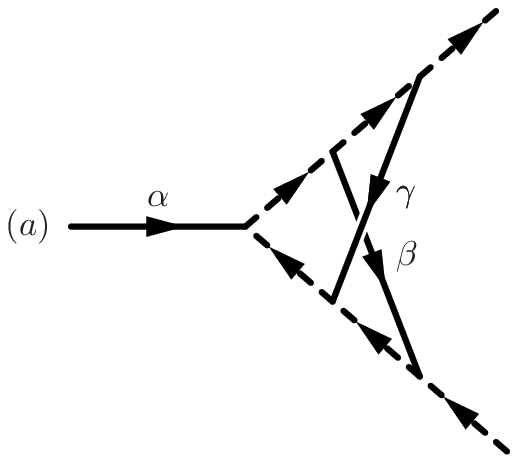}\hspace{0.5cm}
\includegraphics[width=30mm]{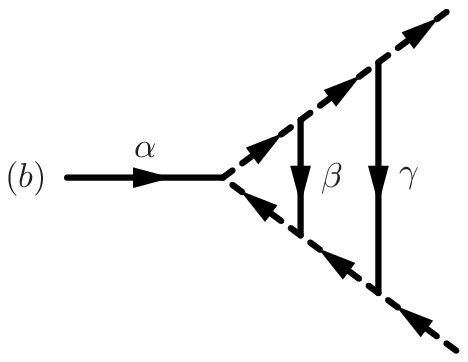}\\
\includegraphics[width=30mm]{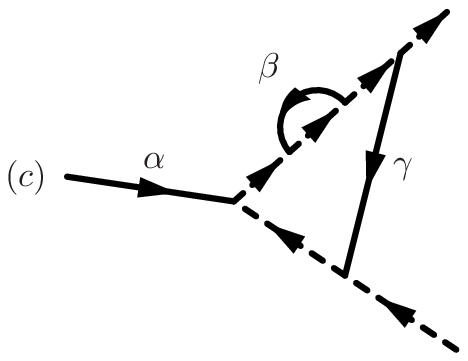}\hspace{0.5cm}
\includegraphics[width=30mm]{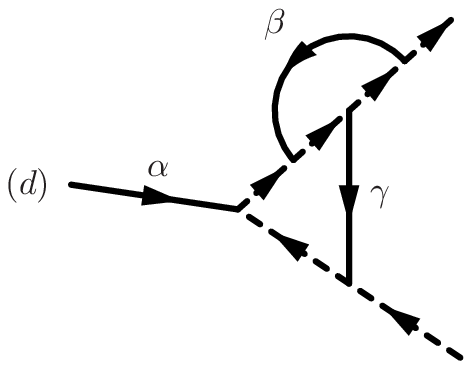}
\caption{The second order contributions to the vertex
$\Gamma_S(\omega_n,\epsilon_n) S_\alpha$.} \label{Fig3}
\end{figure}

The second order corrections to the self energy $\Sigma$ are shown
in Figs.~\ref{Fig2}(a) and \ref{Fig2}(b) and given respectively as
\begin{eqnarray}
\Sigma^{(2)}(\epsilon_n) &=& J^4\mathbf{S}^2_0\mathbf{S}^2_1
T^2\sum_{\omega_s,\omega_m}
D(\omega_s)D(\omega_m)G(\omega_s+\epsilon_n)\notag \\
&&\hspace{1.3cm}\times G(\omega_s+\omega_m+\epsilon_n)
G(\omega_m+\epsilon_n)\notag\\
&+& J^4\mathbf{S}^2_0\mathbf{S}^2_0 T^2\sum_{\omega_s,\omega_m}
D(\omega_s)D(\omega_m)G^2(\omega_s+\epsilon_n)\notag \\
&&\hspace{1.3cm}\times G(\omega_s+\omega_m+\epsilon_n).
\label{sigma2}
\end{eqnarray}
At $T=0$ we find
\begin{eqnarray}
\Sigma^{(2)}(E) =&-&\frac{g^2_0}{4}\mathbf{S}^2_0\mathbf{S}^2_1 E
\left
(\ln^2\frac{E_0}{|E|}+\ln\frac{E_0}{|E|}\right )\notag\\
&+&\frac{g_0^2}{4}\mathbf{S}^2_0\mathbf{S}^2_0 E \left
(\frac{1}{2}\ln^2\frac{E_0}{|E|}+\ln\frac{E_0}{|E|}\right
).\label{Fsigma2}
\end{eqnarray}
The second order corrections to the vertex
$\Gamma_S(\omega_n,\epsilon_n)$ are presented in
Figs.~\ref{Fig3}(a)-(d). They are as follows
\begin{eqnarray}
\Gamma_s^{(2)} &=& J^4\mathbf{S}^2_1\mathbf{S}^2_2
T^2\sum_{\omega_s,\omega_m} D(\omega_s)G(\omega_s+\epsilon_n)
G(\omega_m+\epsilon_n)\notag \\&& \times D(\omega_m)
G(\omega_s+\omega_m+\epsilon_n)
G(\omega_s+\omega_m+\epsilon_n+\omega_n) \notag
\\
&+& J^{2}\mathbf{S}^2_1 T\sum_{\omega_s}
D(\omega_s)G(\omega_s+\epsilon_n)
\Gamma^{(1)}_S(\omega_n,\omega_s+\epsilon_n)\notag \\
&&\hspace{1cm}\times  G(\omega_s+\epsilon_n+\omega_n)\notag
\\
&+& J^{2}\mathbf{S}^2_1 T\sum_{\omega_s}
[D(\omega_s)+D(\omega_s-\omega_n)] G(\omega_s+\epsilon_n)\notag \\
&&\hspace{1cm}\times
G^2(\omega_s+\epsilon_n+\omega_n)\Sigma^{(1)}(\omega_s+\epsilon_n+\omega_n)\notag
\\
&+& J^{4}\mathbf{S}^2_1\mathbf{S}^2_1 T^2\sum_{\omega_s,\omega_m}
D(\omega_s)[D(\omega_m)+D(\omega_m-\omega_n)]\notag \\
&&\hspace{1cm}\times
G(\omega_s+\epsilon_n)G(\omega_s+\omega_m+\epsilon_n+\omega_n)
\notag \\
&&\hspace{1cm}\times G(\omega_s+\epsilon_n+\omega_n)
G(\omega_m+\epsilon_n+\omega_n). \label{vertex2}
\end{eqnarray}
Evaluating Eqs.~\eqref{vertex2} at $T=0$ we find
\begin{eqnarray}
\Gamma_s^{(2)}(0,E) &=&
\frac{g_0^2}{4}\mathbf{S}^2_1\mathbf{S}^2_2
\ln\frac{E_0}{|E|}\notag\\
&+& \frac{g_0^2}{4}\mathbf{S}^2_1\mathbf{S}^2_1
\left (\frac{1}{2}\ln^2\frac{E_0}{|E|}-2 \ln\frac{E_0}{|E|}\right ) \notag\\
&-& \frac{g_0^2}{2}\mathbf{S}^2_0\mathbf{S}^2_1\left
(\frac{1}{2}\ln^2\frac{E_0}{|E|}-\ln\frac{E_0}{|E|}\right )
\notag\\
&+& \frac{g_0^2}{2}\mathbf{S}^2_1\mathbf{S}^2_1\left
(\frac{1}{2}\ln^2\frac{E_0}{|E|}-\ln\frac{E_0}{|E|}\right ).
\label{Fvertex2}
\end{eqnarray}

\subsection{Renormalization group equation for $g$}

Combining together Eqs.~\eqref{Fsigma1}, \eqref{Fvertex1},
\eqref{Fsigma2} and \eqref{Fvertex2} we obtain the following
result
\begin{equation}
\frac{1}{g(E)} =\frac{1}{g_0}+ \ln \frac{E_0}{|E|} + \alpha_S
g_0\ln \frac{E_0}{|E|}, \label{WCginv}
\end{equation}
where
\begin{equation}
\alpha_S=\frac{1}{2}[(S(S+1)-1)^2+1].\label{alphaSdef}
\end{equation}
We emphasize that $\ln^2E_0/|E|$ terms are cancelled out in the
final expression for $1/g(E)$. Following the standard
procedure~\cite{IzyumovSkryabin}, i.e., taking derivative of
$1/g(E)$ with respect to $\xi=\ln |E|/E_0$ and then substituting
$g(E)$ for $g_0$ in the right hand side of Eq.~\eqref{WCginv}, we
obtain the following renormalization group equation
\begin{equation}
\frac{d g}{d\xi} = g^2 + \alpha_S g^3+\mathcal{O}(g^4).
\label{RGWC}
\end{equation}
Equation~\eqref{RGWC} describes the renormalization of the
invariant charge $g$ in the weak coupling limit. The comparison of
two first terms in the right hand side of Eq.~\eqref{RGWC} results
in the inequality $g\ll \alpha_S^{-1}$ that determines the weak
coupling limit. In virtue of the result $\alpha_2=13$ we find the
significant reduction of the weak coupling region from a naive
estimate $g\ll 1$ already for the impurity spin $S=2$.

We mention that evaluation of Eqs.~\eqref{sigma1},
\eqref{vertex1}, \eqref{sigma2} and \eqref{vertex2} at finite
temperatures but $E=0$ results in the same equation as
Eq.~\eqref{RGWC} for the renormalization of $g$ with temperature
but now with $\xi=\ln T/E_0$. This fact is in agreement with
natural expectations of setting the energy scale by temperature
$T$ in this case.

\section{Renormalization group analysis in the strong coupling limit\label{RGS}}

In the strong coupling limit in which $g$ is assumed to be large
it is convenient to use the Holstein-Primakoff representation for
the impurity spin $\mathbf{S}$~\cite{HP}
\begin{gather}
S^+ = a^\dag \sqrt{2S- a^\dag a},\qquad  S^-=\sqrt{2S-a^\dag a}\,
a,\notag \\
S^z=-S+a^\dag a.\label{HPR}
\end{gather}
Here $a^\dag$ and $a$ denotes bosonic creation and annihilation
operators. The effective action~\eqref{Seff} becomes
\begin{gather}
\mathcal{S}_\textrm{eff}= \frac{J^2}{2}\int_0^\beta d\tau_1
d\tau_2 D(\tau_1-\tau_2) \mathbf{S}(\tau_1) \mathbf{S}(\tau_2) \label{SeffHP}\\
+\int_0^\beta d\tau a^\dag (\tau)
\partial_\tau a(\tau).\notag
\end{gather}
The Holstein-Primakoff representation~\eqref{HPR} has the enlarged
Hilbert space as compared with one for the spin $\mathbf{S}$. This
fact becomes immaterial for $g_0 S\gg 1$ such that we can perform
expansion of the square roots in Eq.~\eqref{HPR} in powers of
$a^\dag a$ and perform standard evaluation of a functional
integral with the effective action~\eqref{SeffHP}.

Expanding the product $\mathbf{S}(\tau_1) \mathbf{S}(\tau_2)$ to
the second order in boson fields $a^\dag$ and $a$ we find the bare
propagator $G(\omega_n)$ as follows
\begin{equation}
G^{-1}(\omega_n) = i\omega_n -\pi g_0 S
|\omega_n|/2.\label{propHP}
\end{equation}
The higher order in $a^\dag$ and $a$ terms result in the
renormalization of the $g$. At zero temperature with a help of the
standard background field procedure we derive the following
one-loop result
\begin{equation}
g(E) = g_0 + \frac{2 g}{S} \int^{E_0}_{-E_0} \frac{d\omega}{2\pi}
G(\omega) \Theta(|\omega|-|E|).
\end{equation}
Here $E$ plays a role of the energy scale that separates the
`fast' and `slow' modes. After evaluation we obtain
\begin{equation}
g(E) = g_0 \left [1+ \frac{4}{\pi^2 S^2 g_0}\left (1-
\frac{1}{1+\pi^2 g^2_0 S^2/4}\right ) \ln \frac{|E|}{E_0}\right
].\label{1loopHP}
\end{equation}
If we consider the impurity spin $\mathbf{S}$ being a classical
vector of the fixed length $S$ then at zero temperature the
effective action~\eqref{SeffHP} is equivalent to the so-called
one-dimensional $O(3)$ model with inverse square interaction. In
this case the brackets in front of $\ln |E|/E_0$ in the right hand
side of Eq.~\eqref{1loopHP} should be substituted by unity due to
the absence of $i\omega_n$ term in the propagator~\eqref{propHP}.
Thus, for the $O(3)$ model each loop results in a factor
$g^{-1}_0S^{-2}$. Additional sub-leading series in powers of
$(g_0S)^{-1}$ appears for the action~\eqref{SeffHP} as one can see
from Eq.~\eqref{1loopHP}. Under assumption $g_0 \gg 2/(\pi S)$ we
neglect the sub-leading terms in Eq.~\eqref{1loopHP}. With the
same accuracy in calculation of the two-loop contribution to the
$g(E)$ we omit the term $i\omega_n$ in the
propagator~\eqref{propHP}. Then, the result coincides with the
two-loop result for the $O(3)$ model with inverse square
interaction found in Ref.~\cite{Zwerger}. We obtain therefore
\begin{equation}
g(E) = g_0 + \frac{4}{\pi^2 S^2} \ln \frac{|E|}{E_0} +
\frac{8}{\pi^4S^4 g_0} \ln \frac{|E|}{E_0}. \label{gEHP}
\end{equation}
Hence, we derive the following renormalization group equation
\begin{equation} \frac{d g}{d\xi} =
 \frac{4}{\pi^2S^2}+\frac{8}{\pi^4S^4g}+\mathcal{O}(g^{-2}). \label{RGSC}
\end{equation}
We mention that Eq.~\eqref{RGSC} coincides with the two-loop
renormalization group equation known for the $O(3)$ model with
inverse square interaction~\cite{Zwerger}. The sub-leading terms
neglected above determines the condition $g\gg 2/(\pi S)$ of
applicability for the strong coupling expansion.

\section{Renormalization of $g$ with temperature\label{RNT}}

At relatively high temperatures $T\gtrsim E_0$ the invariant
charge $g$ equals $g_0$. As temperature is lowered the $g$ starts
to be renormalized. In the weak coupling regime
$g\ll\alpha_S^{-1}$ the renormalization of $g$ with $T$ is
determined by Eq.~\eqref{RGWC}. If for a moment we restrict
ourselves to the first term in the right hand side of
Eq.~\eqref{RGWC} we find
\begin{equation}
g(T) = \frac{g_0}{1+g_0 \ln E_0/T}. \label{LM}
\end{equation}
This expression has been obtained originally by Larkin and
Melnikov with a help of summation over the parquet
diagrams~\cite{LarkinMelnikov}. Taking into account the second
term in the right hand side of Eq.~\eqref{RGWC} we obtain ($T\ll
T_0$, $g_0\ll\alpha_S^{-1}$)
\begin{equation}
g(T) = \left [
\ln\left(\frac{T_0}{T}\ln^{\alpha_S}\frac{T_0}{T}\right )\right
]^{-1},\quad T_0 = g_0^{\alpha_S} e^{1/g_0}E_0 .\label{LM2}
\end{equation}
Equation~\eqref{LM2} demonstrates that for $T\ll T_0$ the $g(T)$
decreases as temperature is lowered faster than predicted by the
lowest order result~\eqref{LM} and faster for larger impurity spin
$S$.

In the strong coupling limit $g \gg 2/(\pi S)$ by solving
Eq.~\eqref{RGSC} we find the following temperature dependence
($T\gg \tilde{T}_0$, $g_0 \gg 2/(\pi S)$)
\begin{gather}
g(T) = \frac{4}{\pi^2 S^2}\ln \left
(\frac{T}{\tilde{T}_0}\ln^{1/2} \frac{T}{\tilde{T}_0}\right ),\label{LM3}\\
\quad \tilde{T}_0 = \frac{\pi S \sqrt{g_0}}{2} e^{-\pi^2 S^2
g_0/4}E_0 .\notag
\end{gather}
The result~\eqref{LM3} indicates that for $T\gg \tilde{T}_0$ the
$g(T)$ lowers with decrease of the temperature slower then
logarithmically and faster for large spin $S$.

Let us now assume the existence of the renormalization group
function $\phi(g)=dg/d\xi$ for all values of $g\geqslant 0$ with
asymptotics given by Eqs.~\eqref{RGWC} and \eqref{RGSC}. We
mention that for the impurity spin $S\leqslant 1$~\cite{Foot} the
weak~\eqref{RGWC} and strong~\eqref{RGSC} coupling asymptotics
have a range in which they both are applicable. In the case of
larger $S$ there exists a broad region $\alpha_S^{-1}\ll g \ll
2/(\pi S)$ where the quantitative behavior of function $\phi(g)$
is unknown. In general, it is natural to expect that the function
$\phi(g)$ is positive for all $g>0$ and has the maximum at some
value $g=g_\textrm{max}$.

Existence of the maximum in the renormalization group function
$\phi(g)$ signals about the presence of a new energy scale in the
problem
\begin{equation}
T_\textrm{max} = E_0 \exp \int_{g_0}^{g_\textrm{max}} \frac{d
g}{\phi(g)}.
\end{equation}
For the system with large value of $g_0$ the energy scale
$T_\textrm{max}$ separates temperature ranges of the strong and
weak coupling regimes.

\section{Susceptibility\label{Suscept}}

Direct measurements of the $g(T)$ in a laboratory is impossible.
Fortunately, the information about behavior of the $g(T)$ can be
extracted from the zero field magnetic susceptibility of the
impurity spin. By polarizing electron spins the impurity acquires
a giant magnetic moment $\mu S$ with $|\mu | = \mu_B
|g_\textrm{imp}- g_e J\nu/(1+F_0^\sigma)|\gg
|g_\textrm{imp}|\mu_B$. Here $\mu_B$ denotes the Bohr magneton,
$g_\textrm{imp}$ and $g_e$ the $g$-factors of impurity and
electrons respectively~\cite{LarkinMelnikov,KorenblitShender}.
Thus in the presence of a small magnetic field $h$ effective
action~\eqref{SeffWC} acquires the standard term
\begin{equation}
\mathcal{S}_h=-\mu h\int_0^\beta d\tau c^\dag_m(\tau)S^z_{m
m^\prime} c_{m^\prime}(\tau).
\end{equation}
The zero field magnetic susceptibility of the impurity can be
found as~\cite{IzyumovSkryabin}
\begin{equation}
\chi(T) = T \lim\limits_{h\to 0}\frac{\partial^2}{\partial h^2}
\ln T\sum_{m,\epsilon_n} G_{mm}(i\epsilon_n).\label{suscdef}
\end{equation}
It is convenient to write the general expression for the
susceptibility as follows
\begin{equation}
\chi(T) = \frac{\mu^2 S(S+1)}{3 T}
\Upsilon(g_0,T/E_0).\label{SuscExact}
\end{equation}
Here $\Upsilon(g_0,T/E_0)$ is some dimensionless function and we
singled out the typical factor of the Curie-Weiss susceptibility
for a free spin in Eq.~\eqref{SuscExact} for convenience. Let us
now perform a change of the temperature scale $T \to \lambda T$
such that~\cite{ZinnJustin}
\begin{equation}
\Upsilon(g_0,T/E_0) = Z(\lambda) \Upsilon(g(\lambda^{-1}),\lambda
T/E_0). \label{Change}
\end{equation}
The conditions $Z(1)=1$ and $g(1) = g_0$ should be obviously
imposed. The right hand side of Eq.~\eqref{Change} should be
independent of an arbitrary scale parameter $\lambda$. Thus we
obtain the following equation for $Z(\lambda)$
\begin{equation}
\frac{d\ln Z(\lambda)}{d\ln \lambda} = - \frac{d\ln
\Upsilon(g(\lambda^{-1}),\frac{\lambda T}{E_0})}{d\ln
\lambda}\equiv \zeta(g(\lambda^{-1})).\label{Zren}
\end{equation}
By solving Eq.~\eqref{Zren} for $\lambda=E_0/T$ and taking into
account that $g(T/E_0)\equiv g(T)$ we find
\begin{equation}
\chi(T) = \frac{\mu^2S(S+1)}{3T F(g_0)} F(g(T)),\label{SuscExact2}
\end{equation}
where we have used the fact that $\Upsilon(g(T),1)=1$ and
\begin{equation}
F(g) = \exp\left (- \int_{g_\textrm{max}}^g du
\frac{\zeta(u)}{\phi(u)}\right ). \label{Fg0}
\end{equation}

In the weak coupling regime we compute
$\Upsilon(g(\frac{1}{\lambda}),\frac{\lambda T}{E_0})$
perturbatively in $g(\lambda^{-1})$  for $\lambda$ of the order of
$E_0/T$. To the lowest order in $g(\lambda^{-1})$ the
pseudofermion Green function in the presence of a small magnetic
field $h$ becomes
\begin{equation}
G^{-1}_{mm^\prime}(\epsilon_n) = \delta_{mm^\prime}\left
[(i\epsilon_n +\eta) \Gamma(\epsilon_n) - h m
\Gamma_S(0,\epsilon_n)\right ].
\end{equation}
Here $\Gamma(\epsilon_n)$ and $\Gamma_S(0,\epsilon_n)$ are given
by Eqs.~\eqref{Fsigma1} and \eqref{Fvertex1}. With a help of
Eq.~\eqref{suscdef} we find
\begin{equation}
\Upsilon(g(\lambda^{-1}),\lambda T/E_0) = 1+g(\lambda^{-1})\ln
\frac{\lambda T}{E_0}.\label{Ups}
\end{equation}
Then Eq.~\eqref{Ups} yields
\begin{equation}
\zeta(g)=-g-\mathcal{O}(g^2)\label{zetaWC}
\end{equation}
and we obtain the function $F(g)$ at $g\ll \alpha_S^{-1}$ as
\begin{equation}
F(g) = \textrm{const}\, g(1+\mathcal{O}(g)).\label{FgWC}
\end{equation}
Equations~\eqref{Ups} and \eqref{FgWC} allow us to derive the
following result for the susceptibility $\chi(T)$ in the weak
coupling regime, $g\ll\alpha_S^{-1}$,
\begin{equation}
\chi(T)=\frac{\mu^2S(S+1) g(T)}{3 T g_0}.\label{chiTW}
\end{equation}
We mention that the result~\eqref{chiTW} with $g(T)$ given by
Eq.~\eqref{LM} was first found by Larkin and
Melnikov~\cite{LarkinMelnikov} with a help of a direct summation
of the parquet diagrams. The derivation presented above is more
general since the true invariant charge $g(T)$ enters into
Eq.~\eqref{chiTW}. With a help of Eq.~\eqref{chiTW} we find the
temperature behavior of the susceptibility $\chi(T)$ at $T\ll T_0$
($g_0\ll\alpha_S^{-1}$) as follows
\begin{equation}
\chi(T)=\frac{\mu^2S(S+1)}{3 T g_0}\left [
\ln\left(\frac{T_0}{T}\ln^{\alpha_S}\frac{T_0}{T}\right )\right
]^{-1}.\label{chiTWF}
\end{equation}

In the strong coupling limit we can perform similar analysis as
above by expansion in powers of $1/g(\lambda^{-1})$ when $\lambda$
is of the order of $E_0/T$. By using the perturbative result
\begin{equation}
\Upsilon(g(\lambda^{-1}),\lambda T/E_0) = 1+\frac{4}{\pi^2 S(S+1)
g(\lambda^{-1})}\ln \frac{\lambda T}{E_0}\label{Ups2}
\end{equation}
we obtain
\begin{equation}
\zeta(g)=-\frac{4}{\pi^2 S(S+1)
g}+\mathcal{O}(g^{-2}).\label{zetaSC}
\end{equation}
Hence, we find
\begin{equation}
F(g) = \textrm{const}\, g^{S/(S+1)} (1+
\mathcal{O}(g^{-1})).\label{FgSC}
\end{equation}
With a help of Eqs~\eqref{Ups2} and \eqref{FgSC} we derive the
following result for the susceptibility $\chi(T)$ in the strong
coupling regime, $g\gg 2/(\pi S)$,
\begin{equation}
\chi(T)=\frac{\mu^2S(S+1)}{3T} \left [\frac{g(T)}{g_0}\right
]^{S/(S+1)} .\label{chiTS}
\end{equation}
By using Eq.~\eqref{LM3}, we obtain the temperature behavior of
the susceptibility at $T\gg \tilde{T}_0$ ($g_0\gg 2/(\pi S)$) as
follows
\begin{equation}
\chi(T)=\frac{\mu^2S(S+1)}{3T g_0^{S/(S+1)}} \left [
\frac{4}{\pi^2 S^2}\ln \left (\frac{T}{\tilde{T}_0}\ln^{1/2}
\frac{T}{\tilde{T}_0}\right ) \right ]^{S/(S+1)}.\label{chiTSF}
\end{equation}
It is worthwhile to mention that for $S\gg 1$ when the impurity
spin becomes classical the function $T\chi(T)$ is proportional to
$g(T)$ in both weak and strong coupling regimes.

\begin{figure}[tbp]
\includegraphics[width=80mm]{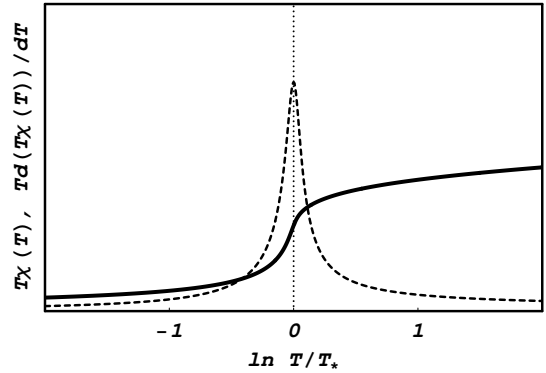}
\caption{Sketch for dependencies of the functions $T \chi(T)$
(solid curve) and $Td(T\chi(T))/dT$ (dashed curve) on $\ln
T/T_\star$. Both curves are obtained by interpolation between the
weak~\eqref{chiTW} and strong~\eqref{chiTS} coupling asymptotics.
See text.} \label{Fig4}
\end{figure}

According to Eqs.~\eqref{chiTW} and \eqref{chiTS} the quantity
$T\chi(T)$ decreases with lowering temperature. The derivative $T
d(T\chi(T))/dT$ should have the maximum at some intermediate
temperature $T_\star$ that determines by the condition
$\zeta^2(g)=\phi(g)\zeta^\prime(g)$. We mention that in general
the $T_\star$ does not coincide with the energy scale
$T_\textrm{max}$. We present sketches for the temperature behavior
of the functions $T\chi(T)$ and $T d(T\chi(T))/dT$ in
Fig.~\ref{Fig4}.

\section{Conclusions\label{Conc}}

In conclusions we performed the renormalization group analysis for
the problem of the magnetic impurity in nearly ferromagnetic Fermi
liquid. We evaluated the second order term in the weak coupling
expansion of the renormalization group equation for the invariant
charge $g$. We derived the two-loop renormalization group equation
for the $g$ in the strong coupling regime. We found the low and
high temperature asymptotics of the magnetic susceptibility of the
impurity spin $\chi(T)$ and showed that the derivative $T
d(T\chi(T))/dT$ has the maximum.

In the discussion above we have neglected the usual Kondo
contributions~\cite{Kondo,IzyumovSkryabin} which involve odd
powers of $J\nu$. However such terms do not contain the
anomalously large coefficients $(1+F_0^\sigma)^{-1}$ as it occurs
in terms of even powers considered above. As it was originally
shown by Larkin and Melnikov~\cite{LarkinMelnikov} the condition
$g\gg|J|\nu$, or equivalently, $|J|\nu\gg 1+F_0^\sigma$ is
sufficient to omit odd in $J\nu$ terms. The results obtained above
are valid therefore only at not too low temperatures $T\gg T_l=E_0
\exp[-c(1+F_0^\sigma)^{-1}]$ where $c=1/(16a^2p_F^2)$. For $c$ of
the order of unity and $1+F_0^\sigma\ll 1$ the temperature $T_l$
can be of the several orders in magnitude less than $E_0$. At the
temperature $T_l$ the giant magnetic moment of the impurity $\mu$
is totally compensated such that the impurity susceptibility
becomes $\chi(T_l)\sim \mu_B^2 S(S+1)/T_l$. At lower temperatures
$T \lesssim T_l$ the usual physics of the multichannel Kondo
problem~\cite{MultiKondo} with the number of channels of the order
of $c(1+F_0^\sigma)^{-1}$ manifests. It is worthwhile to mention
that the presence of the paramagnons complicates the subject as
compared with the standard multichannel Kondo problem at
$T\lesssim T_l$~\cite{Maebashi}.

Also we mention that the temperature dependence of part of the
resistance related with scattering on the magnetic impurities
should have the same features as quantity $T\chi(T)$ for not too
low temperatures $T\gg T_l$~\cite{LarkinMelnikov}.

Finally, we emphasize that new detailed experimental
investigations of the zero field magnetic susceptibility of
impurity as well as the resistance in nearly ferromagnetic Fermi
liquid are necessary to test the predictions made in the paper.

\begin{acknowledgement}
It is a pleasure for me to thank N.\,M.\,Chtchelkatchev for
fruitful discussions during the course of this work and
M.\,V.\,Feigelman for the interest to this research and for
critical remarks. I thank S.\,E.\,Korshunov and M.\,A.\,Skvortsov
for useful discussions. The financial support from the Russian
Ministry of Education and Science, Council for Grants of the
President of Russian Federation, Russian Science Support
Foundation, and Dutch Science Foundations \textit{FOM} and
\textit{NWO} is acknowledged.
\end{acknowledgement}


\begin{thebibliography}{30}
\bibitem{Footnote2} In general, there are higher order terms in
the series expansion of the
$\mathcal{D}^{-1}(\mathbf{k},\omega_n)$ in powers of
$|\omega_n|/(kv_F)$. If we take the term proportional
$\omega_n^2/(kv_F)^2$ into account then we find that
$\mathrm{D}(z) \approx 1 + (2/\pi+\pi/2)(1+F_0^\sigma)z$ for $z\ll
1$ and $\mathrm{D}(z) \approx 4\pi /(3^{3/2}z^{2/3}) +
(4/\pi)(1+F_0^\sigma) \ln z$ at $z\gg 1$
\end{thebibliography}
\end{document}